\documentclass[aps,prd,nofootinbib,twocolumn,floatfix,eqsecnum]{revtex4}
\usepackage{dcolumn}
\usepackage{times,mathptm}
\usepackage{subfigure}
\usepackage{graphicx,psfrag,pifont,color,setspace}
\usepackage{amsfonts,amsthm,bm,bbm,amsmath,amssymb,marvosym} 
\usepackage[automark]{scrpage2}
\newcommand{\beq}{\begin{equation}}
\newcommand{\eeq}{\end{equation}}
\newcommand{\bea}{\begin{eqnarray}}
\newcommand{\eea}{\end{eqnarray}}

\newcommand{\C}{{\cal C}}

\newcommand{\G}{{\cal G}}

\renewcommand{\d}{\delta}
\renewcommand{\l}{\lambda}

\renewcommand{\b}{\beta}

\renewcommand{\k}{\kappa}

\newcommand{\g}{\texttt{g}}

\newcommand{\m}{\mu}
\renewcommand{\r}{\rho}

\newcommand{\bx}{{\mathbf{x}}}
\newcommand{\by}{{\mathbf{y}}}

\newcommand{\s}{\sigma}

\newcommand{\E}{{\cal E}}

\renewcommand{\th}{\theta}

\newcommand{\oh}{{\textstyle{\frac{1}{2}}}}

\newcommand{\oq}{{\textstyle{\frac{1}{4}}}}

\newcommand{\dg}{\dagger}
\newcommand{\non}{\nonumber}

\newcommand{\rf}[1]{(\ref{#1})}
\newcommand{\ra}{\rightarrow}
\newcommand{\pa}{\partial}

\begin{document}

\title{Gauge Orbits and the Coulomb Potential}

\author{J. Greensite}
\affiliation{Physics and Astronomy Dept., San Francisco State
University, San Francisco, CA~94132, USA}

\date{\today}
\begin{abstract}

    If the color Coulomb potential is confining, then the Coulomb field energy of an isolated color charge is infinite
on an infinite lattice, even if the usual UV divergence is lattice regulated.   A simple criterion for Coulomb confinement
is that the expectation value of timelike link variables vanishes in Coulomb gauge, but it is unclear how this criterion
is related to the spectrum of the corresponding Faddeev-Popov operator, which can be used to 
formulate a quite different criterion for Coulomb confinement.    The purpose of this article is to connect the two 
seemingly different Coulomb confinement criteria, and explain the geometrical basis of the connection.  

\end{abstract}

\maketitle

\section{\label{intro}Introduction}

   The Coulomb potential in non-abelian gauge theories is of interest for several reasons.   First of all,
since a confining Coulomb potential is a necessary (though not sufficient) condition for having a confining
static quark potential \cite{Dan}, an understanding of the former type of potential could be helpful in understanding
the latter, at least in the Casimir-scaling regime.  Secondly, the Coulomb potential may be useful in various 
hadron phenomenology and spectrum calculations, perhaps along the lines suggested by Szczepaniak and 
co-workers \cite{Adam}.  Finally, the confining Coulomb potential is an important ingredient in the ``gluon-chain" 
model \cite{gchain}, which is a theory of the formation of color electric flux tubes (for a recent development, see \cite{new}). 

    If there is a confining Coulomb potential, then the Coulomb energy of an isolated color charge in an infinite volume
must be infinite, even with a lattice regulation of the usual ultraviolet divergence, and this condition can be expressed in
two very different ways.  The first, derived in ref.\ \cite{GOZ1}, is obtained from the expectation value of the non-local term in the Coulomb gauge Hamiltonian, and is a condition on the density of near-zero modes of the Faddeev-Popov operator.  The second is
the criterion that the expectation value of timelike link variables vanishes in Coulomb gauge \cite{GOZ2}.  These two criteria
are so different that they appear to have, at best, only a very indirect relationship.  In this article I will show how one criterion depends on the other,  and discuss the relevant properties of gauge orbits which underlie this dependency.

    It is important to first understand where these two different Coulomb confinement criteria come from.   Consider the physical state
\beq
         \Psi^a_q = q^a(\bx) \Psi_0 
\eeq
in Coulomb gauge, where $q^a$ is a heavy quark operator, and $\Psi_0$ is the ground state.  Let
\beq
          T = \exp[-(H-E_0)a]
\eeq
represent the lattice theory transfer matrix, divided by a factor of $\exp(-E_0 a)$, where $E_0$ is the vacuum energy and $a$ is the
lattice spacing.
Then
\bea
\exp[-E_{self}] &\equiv& \langle \Psi^a_q |T|\Psi^a_q \rangle 
\non  \\
&=& \langle \text{Tr}[U_0(\bx,t)] \rangle
\non \\
   &\ra& 0
\eea
where $E_{self}$ is the Coulomb energy, in lattice units, of the isolated heavy quark state, and the last line holds if this self-energy is infinite in an infinite volume (with the UV divergence controlled by the finite lattice spacing).  This means that Coulomb confinement, on the lattice,
is equivalent to having a vanishing expectation value for the trace of timelike link variables.  It was noted long ago that 
Coulomb gauge does not fix the gauge completely; there is a remnant gauge symmetry which depends on time but is homogenous in space:
\bea
             U_i(\bx,t) &\ra& g(t) U_i(\bx,t) g^\dg(t)
\non \\
             U_0(\bx,t) &\ra& g(t) U_0(\bx,t) g^\dg(t+1)
\label{remnant}
\eea
If remnant gauge symmetry is unbroken, then $\langle \text{Tr}[U_0]\rangle=0$ \cite{Parisi}, and the Coulomb energy of an isolated 
color charge is infinite \cite{GOZ2}.

    On the other hand, the Coulomb energy of an isolated charge can alternatively be expressed in terms of the
inverse Faddeev-Popov operator:   
\beq
E_{self} = {\g^2 C_F \over N^2 -1} \Bigl \langle (M^{-1} (-\nabla^2) M^{-1} )_{\bx \bx}^{aa} \Bigl \rangle   
\eeq   
where $C_F$ is the quadratic Casimir of the fundamental representation of the SU(N) gauge group, $\g$ is the gauge coupling, and
\beq
             M^{ab}_{\bx \by}[A] =  - (\nabla \cdot D^{ab})_{\bx} \d(\bx - \by)
\eeq
is the Faddeev-Popov (F-P) operator, with $D_k^{ab}$ the covariant derivative.  Let $\{\l_n,\phi_n^a(x)\}$ denote the 
set of eigenvalues and eigenstates of the F-P operator
\beq
        M^{ab}_{\bx \by} \phi^b_n(\by) = \l_n \phi^a_n(\bx)
\eeq
Then it is fairly straightforward to show that in an infinite volume, where the spectrum of $M$ is continuous,
we have \cite{GOZ1}
\beq
E_{self} \propto \g^2 \int_0^{\l_{max}} d\l ~ \left\langle \r(\l) {(\phi_\l|(-\nabla^2)|\phi_\l) \over \l^2}\right \rangle
\label{EC}
\eeq
where $\r(\l)$ is the density of eigenvalues, normalized to unity.   Coulomb confinement ($E_{self}=\infty$) requires that integral in \rf{EC} diverges from a singularity in the integrand at the lower limit, due to the near-zero eigenmodes.  The existence of such eigenmodes implies that the Coulomb gauge lattice configuration, 
in an infinite volume, lies at the Gribov horizon, in accordance with the Gribov-Zwanziger scenario \cite{Dan1}.   However, while proximity
to the horizon is certainly a necessary condition for Coulomb confinement, it is not sufficient.  The density and behavior of near-zero 
eigenmodes must also be such that the integral in eq.\ \rf{EC} is divergent.

   We therefore have two, apparently quite distinct, criteria for Coulomb confinement:
\bea
          &\text{\bf A)}& \qquad   \lim_{\l \ra 0}   {\left\langle \r(\l) (\phi_\l|(-\nabla^2)|\phi_\l) \right \rangle\over \l}  >  0
\label{A}
\\ \non \\
           &\text{\bf B)}& \qquad    \langle \text{Tr}[U_0(\bx,t)] \rangle = 0
\label{B}
\eea
These conditions look very different.  They are, in fact, associated with two different ways of computing the Coulomb
potential.    The first is to calculate the potential
\beq 
           V_C(R) = - \g^2 C_F {1\over N^2-1} \Bigl \langle [M^{-1}(-\nabla^2)M^{-1}]^{aa}_{\bx \by} \Bigr \rangle  
\label{vc}
\eeq
directly, on the lattice, and a number of authors have followed this approach \cite{direct}.  The published results
provide data for the Coulomb potential in momentum space, which appears to go as $1/k^4$ at small $k$.   A second, and
computationally much simpler method was suggested in ref.\ \cite{GO}, and only requires computing the
correlators of timelike link variables, in Coulomb gauge, at equal times.  Define
\beq
         \Psi_{q\overline{q}} = \overline{q}^a(0) q^a(R) \Psi_0
\eeq
and let $a_t = \xi a$ be the lattice spacing in the time direction.  Then
\bea
          \exp[-\xi \E(R)] &\equiv&\langle \Psi_{q\overline{q}}| T |\Psi_{q\overline{q}} \rangle 
\non \\
                &=& \langle \Psi_{q\overline{q}} |e^{- (H-E_0) a_t} |\Psi_{q\overline{q}} \rangle 
\non \\
                &=&  \langle \text{Tr}[U_0^\dg(\bx,t) U_0(\by,t)] \rangle
\eea
so that the Coulomb energy $\E(R)$ associated with a static $q\overline{q}$ state, in units of the
spatial lattice spacing $a$, is given by the timelike link-link correlator
\bea
              \E(R) &=& -{1\over \xi} \log\Bigl[\langle \text{Tr}[U_0^\dg(\bx,t) U_0(\by,t)] \rangle\Bigr]          
\non \\
                        &=& V_C(R) + \text{const}
\label{my_way}
\eea
where $R=|\bx-\by|$.  This approach has been followed (at $\xi=1$)
 in refs.\ \cite{GO,japan}.  The Coulomb string tension $\s_C$ is extracted from the
exponential falloff of the timelike link-link correlator
\beq
      \langle \text{Tr}[U_0^\dg(\bx,t) U_0(\by,t)] \rangle \sim e^{-\xi \s_C R}
\label{UU}
\eeq
and of course this exponential falloff is only possible if condition {\bf B} (eq.\ \rf{B} above) is satisfied.
The interesting question is \emph{why} $\langle \text{Tr}[U_0]\rangle = 0$, and how this condition is related
to the Coulomb confinement criterion {\bf A} (eq.\ \rf{A}), which is formulated in terms of the spectrum of the
F-P operator.

\section{Gauge orbits and their near-tangential intersections}     
     
    The suggestion I will make here is that the condition $\langle \text{Tr}[U_0]\rangle = 0$, and the existence of
a finite correlation length among timelike link variables, is associated with the way in which a typical gauge orbit intersects 
the submanifold of gauge fields satisfying the Coulomb gauge condition, and that this in turn is related to the density
of near-zero F-P eigenmodes.  In continuum notation, let $A_k^a(\bx)$
represent a gauge field at some fixed time $t$ (spatial index $k=1,2,3$), and
\beq
       F^a_\bx[A] \equiv \nabla \cdot {\bf A}^a(\bx) = 0  
\label{F}
\eeq
is the Coulomb gauge condition.  The Fadeev-Popov (F-P) operator $M$ is given by 
\bea
       M^{ab}_{xy}[A] &=& - \g \left( {\d \over \d \theta^b(\by)} F^a_\bx[g\circ A] \right)_{|_{\theta=0}}
\non \\
          &=& - \partial_k D^{ab}_k \d(\bx-\by)
\label{M}
\eea
where
\beq
           g(\bx) = \exp[i \theta^a(\bx) T_a]
\eeq
is a gauge transformation.  

    Let $\C$ represent the hypersurface, in the space of gauge fields $A_k(\bx)$ in $D=3$ dimensions, satisfying the gauge condition
condition \rf{F}.  For a given $A' \in \C$, the corresponding $\{\phi_n\}$ can be thought of as a set of orthonormal unit vectors which span the tangent space, at the point $g(\bx)=\mathbbm{1}$, of the space of all gauge transformations.  Since the gauge orbit $O[A']$ consists of the set of all configurations $g\circ A'$, the  $\{\phi_n\}$ also map to a set of directions spanning the tangent space of the gauge orbit, at the point $A' \in \C$.  

     It was found in ref.\ \cite{GOZ1} that for typical (i.e.\ Monte-Carlo generated) lattices, transformed
to Coulomb gauge, there is a very large number of near-zero modes with $\l_n  \ll 1$, at least as compared to the corresponding spectrum
of the free-field operator $-\d^{ab} \nabla^2$, and this number grows linearly with the lattice volume. 
Near zero-modes have a geometrical intepretation: these are ``flat" directions on the gauge orbit at point $A$, which are nearly tangential to $\C$. 
Said in another way:  a great many directions on $\C$ run nearly parallel to the gauge orbits.   

\begin{figure}[t!]
\centerline{\scalebox{0.6}{\includegraphics{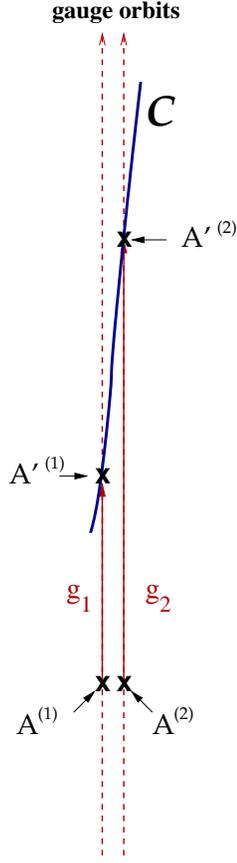}}}
\caption{If the gauge-fixing hypersurface $\C$ defined by $\nabla \cdot A = 0$ is almost tangential to
many directions on a typical gauge orbit, then the gauge transformations $g_1$ and $g_2$ which take two
nearby configurations $A^{(1)}$ and $A^{(2)}$ into Coulomb gauge may be very different.}
\label{orbits}
\end{figure}

    To understand the implications of this fact, let us consider two points  $A_k^a(\bx)$ and  $A_k^a(\bx) + \d A_k^a(\bx)$ with an infinitesimal separation $\d A$ in the space of all three-dimensional gauge field configurations (a time slice of the $D=4$ configurations, ignoring the $A_0$ component).  Let $g_1$ and $g_2$ be the gauge transformations which bring $A^{(1)}=A$ and $A^{(2)}=A+\d A$, respectively, 
onto $\C$ (Fig.\ \ref{orbits}) .  Of course,
$g_1$ and $g_2$ are not unique because of Gribov copies, and also because of the remnant global gauge symmetry \rf{remnant}
allowed by Coulomb gauge.
However, for a given $g_1$ the ambiguity in $g_2$ can be eliminated by the requirement that $g_2$ brings $g_2\circ (A+\d A)$ as close as possible to $g_1\circ A$.  Then for $\d A$ infinitesimal, the deviation of $g_2$ from $g_1$ must also be
infinitesimal, and we can write
\beq
              g_2(\bx) = \exp[i\d \th^a(\bx) T_a] g_1(\bx)
\eeq
where $T_a=\oh \s_a$ for the SU(2) group used below.
The deviation $\d \th$ is determined from the condition that 
\beq
               F^a_x\Bigl[ g_2 \circ (A+\d A) \Bigr] = 0
\label{Fg2}
\eeq
Expanding this condition to first order in a functional Taylor series, and taking account of $F[g_1\circ A] = 0$, we have
\beq
               {\d F^a_\bx \over \d (\d\th^b(\by))} \d \th^b(\by) + {\d F^a_\bx \over \d(\d A_k^b(\by))} \d A_k^b(\by) = 0
\label{taylor}
\eeq
where repeated indices are summed, repeated coordinates ($\by$) are integrated, and the functional derivatives are evaluated at
$\d \th = \d A = 0$.  Then, from eq.\ \rf{M},
\beq
                  M^{ab}_{\bx \by}[g_1 \circ A] \d \th^b(\by) = \g \d f^a(\bx)
\label{solve}
\eeq
where
\bea
\d f^a(\bx) &\equiv& {\d F^a_\bx \over \d A_k^b(\by)} \d A_k^b(\by)
\non \\
&=& \pa_k \Bigl\{ \oh \text{Tr}[g_1(\bx) \s_b g_1^\dg(\bx) \s_a] \d A^b_k(\bx) \Bigr\}
\eea
Now let $\{\phi_n\}$ be the eigenstates of the F-P operator $M$ at
$g_1\circ A$, and expand \footnote{The $\phi_n$ are real-valued, so the complex-conjugation symbol in eqs.\ \rf{expand},
\rf{fl}, and \rf{avtheta} below is superfluous.  It is retained nonetheless, to indicate that the eigenstate would be a bra vector in bra-ket notation.}
\bea
            \d \th^a(\bx) &=& \sum_{n>3} \d \th_n ~\phi_n^a(\bx)
\non \\
            \d  f^a(\bx) &=& \sum_{n>3} \d f_n ~\phi_n^a(\bx)
\non \\
             M^{ab}_{\bx \by}  &=&  \sum_{n>3} \l_n \phi^a_n(\bx) \phi^{b*}_n(\by)
\label{expand}
\eea
The restriction to $n>3$ in the above summations has to do with remnant global gauge symmetry in Coulomb gauge.  For the SU(2) gauge group, the F-P operator has three exact zero modes ($\l_{1-3}=0$) on a finite volume with periodic boundary conditions, corresponding to the fact that if a configuration $A$ satisfies the Coulomb gauge condition, so does $g\circ A$ for a spatially independent (i.e.\ global) gauge transformation.  For this reason, the coefficients $\d \th_{1-3}$ are not determined by the condition \rf{Fg2}, and can be set to anything we choose; in particular 
they can be set to zero.  Note also that $\d f_{1-3}=0$.  This is because the three zero modes are
constant in space, while $\d f(\bx)$ is a total derivative, so the inner product of $(\phi_n|\d f)$ vanishes for $n=1,2,3$.  

     Substituting the expansions \rf{expand} into \rf{solve}, we find for all $n>3$ that
     \beq
            \d \th_n = \g {\d f_n \over \l_n}
\eeq
This equation takes us to the crux of the matter.
The $\d f_n$ coefficients depend on $\d A_k^a(\bx)$, which is small but otherwise arbitrary, so near-zero $\l_n$ correspond in general
to very large $\d \th_n$.    If there are only a few near-zero eigenmodes,
then the few large $\d \th_n$ may not contribute very much to $\d \th(\bx)$.  On the other hand, if there are a large number of
near-zero eigenmodes, then $\d \th(\bx)$ may also be large, and the deviation between $A^{(1)}$ and $A^{(2)}$ 
will be greatly magnified upon gauge-fixing to $\C$, as illustrated in Fig.\ \ref{orbits}.  

    The quantity to consider is the mean-square value of $\d \th(\bx)$, which is given by
\bea
          \overline{|\d \th|}^2 &=& {1\over V_3} \int d^3x ~ \d \th^a(\bx) \d \th^a(\bx) 
\non \\
                &=&  {1 \over V_3} \sum_n \d \th_n^2
\non \\
                &\stackrel{V_3 \ra \infty}{\longrightarrow}& \int d\l ~ \r(\l) \d \th_\l^2
\non \\
                &=& \g^2 \int d\l ~ {\r(\l) [\d f(\l)]^2 \over \l^2 }
\label{dth}
\eea
where
\beq
\d f(\l) = \int d^3x ~ \phi^{a*}_\l(\bx) \pa_k \text{Tr}[g_1 \d A_k g_1^\dg \s_a]
\label{fl}
\eeq
and $V_3$ is the three-volume of a time-slice.
Let us consider the magnitude of $\overline{|\d \th|}^2$ for a ``typical" $\d A$.  To derive this quantity, we need to average over
the $\d A$ with some reasonable, gauge-invariant probability measure.\footnote{Note that $\d A$ transforms homogeneously,
i.e.\ $\d A \ra g \d A g^\dg$ under a gauge transformation.}  The simplest is a gaussian measure
\bea
 \lefteqn{\langle Q[\d A] \rangle_{gauss} } 
 \non \\
    &=& { \int D \d A ~ Q[\d A] \exp\left[-{1\over 2\epsilon} \int d^3x ~ \d A^a_k(\bx) \d A^a_k(\bx) \right]
                       \over  \int D \d A ~  \exp\left[-{1\over 2\epsilon} \int d^3x ~ \d A^a_k(\bx) \d A^a_k(\bx) \right] }
\non \\
\eea
where $\epsilon$ is an infinitesimal constant.  In this measure
\beq
            \langle \d A_i^a(\bx) \d A_j^b(\by)  \rangle_{gauss}  =  \epsilon \d_{ij} \d^{ab} \d^3(\bx-\by)
\eeq
Then
\bea
 \lefteqn{\langle \overline{|\d \th|}^2 \rangle_{gauss} }
\non \\   
     &=&      \g^2   \int d\l ~ {\r(\l) \over \l^2 }  \int d^3 x d^3 y \Bigl\langle \oq (\pa^i_x \phi^{a*}_\l(\bx)) 
             \text{Tr}[g_1(\bx)\s_b g_1^\dg(\bx) \s_a] 
\non \\
     & & \qquad \d A^b_i(\bx) \text{Tr}[g_1(\by)\s_c g_1^\dg(\by) \s_d] \d A^c_j(\by) (\pa^j_y \phi^d_\l(\by)) \Bigr\rangle_{gauss}
\non \\
     &=& \epsilon \g^2 \int d\l ~ {\r(\l) (\phi_\l | (-\nabla^2) | \phi_\l ) \over \l^2 } 
\label{avtheta}
\eea
The above expression still depends on the initial choice of $A$, since the F-P eigenmodes are evaluated at $g_1\circ A$, 
but if we now take the vacuum expectation value, then the 
integral is precisely the same as the integral that appears in the expression for the Coulomb energy 
of an isolated charge, shown in eq.\ \rf{EC}.\footnote{We make use here of the fact that $\langle Q[g_1\circ A] \rangle$, evaluated without
gauge-fixing, is the same as $\langle Q[A] \rangle$ evaluated in Coulomb gauge, as first pointed out by Mandula and Ogilivie \cite{Mandula}.}

      There are now two possible scenarios, depending on whether or not the Coulomb confinement criterion {\bf A} (eq.\ \rf{A}) is satisfied.
Let
\beq
        G(\bx) \equiv \exp[i\d \th^a(\bx) T_a] = g_2(\bx) g_1^\dg(\bx) 
\eeq
serve to quantify the deviation between gauge transformations $g_1$ and $g_2$. 
Making use of the global remnant symmetry \rf{remnant}, it is always possible to set $G(\bx)=\mathbbm{1}$ at one particular site $\bx=\bx_0$.
If there is no Coulomb confinement, so the integral in \rf{avtheta} is finite, then $\d \th(\bx)$ is everywhere small for sufficiently small $\d A$.  This
means that $G(\bx) \approx \mathbbm{1}$ and $g_2(\bx) \approx g_1(\bx)$ everywhere in space.  On the other hand, if the Coulomb confinement  condition {\bf A} is satisfied, a quite different scenario is possible.  It can then happen that no matter how small the magnitude of $\d A$,
the non-compact variable $\d \th(\bx)$ becomes large at sufficiently large $|\bx-\bx_0|$, and $\d \th$ is a random variable because $\d A$ is a random
variable.  In this case, the most likely behavior is that as $\bx$ varies, $G(\bx)$ wanders over the entire group manifold, averaging to zero
in an infinite volume.   Assuming that  $G(\bx)$ averages to zero, a better measure 
of the $g_{1,2}$ deviation is provided by the correlation length among the $G(\bx)$, extracted from the correlator
\beq
             D(R) = {1\over V_3} \int d^3 x  \oh \text{Tr}[G(\bx) G^\dg(\bx + {\bf R})]  
\label{DR1}
\eeq
rather than $G(\bx)$ itself.   In particular, if $D(R)$ goes as
\beq
              D(R) \sim e^{-\m R} 
\label{DR2}
\eeq
at large $R$, then there is a finite correlation length $l_g = \m^{-1}$, and $\langle G \rangle = 0$.  As $\d A \ra 0$ (so $g_2 \ra g_1$), we would
expect this correlation length to  go to infinity, i.e.\ $\m \ra 0$.  Confinement criterion {\bf A} is a \emph{necessary} condition for this kind of behavior,
but since eq.\ \rf{taylor} is not necessarily valid for $\d \th \sim O(1)$, we must resort, in this case, to a numerical investigation.
     
       To summarize:  Near-zero modes of the F-P operator correspond to directions in the gauge orbit $O[A]$ which are nearly tangential
to the gauge-fixed hypersurface $\C$, at the point where $O[A]$ intersects $C$.   Large numbers of tangential 
directions should have the following consequence:  if $A$ is a typical gauge field on a time slice, and $A+\d A$ is a nearby configuration on
the same timeslice, then the gauge transformations $g_1$ and $g_2$ which take $A$ and $A+\d A$ into Coulomb gauge will be wildly different,
even for relatively small $\d A$.   More concretely, in theories where the Coulomb energy of an isolated color charge is infinite, eq.\ 
\rf{avtheta} suggests that $G(\bx) = g_2(\bx) g^\dg_1(\bx)$ is a random variable ($\langle G(\bx) \rangle = 0$) with a finite $GG$ correlation length.   
On the lattice it is possible to test this possibility, by calculating $D(R)$ numerically.

\section{Numerical Results}

\begin{figure}[t!]
\centerline{\scalebox{0.7}{\includegraphics{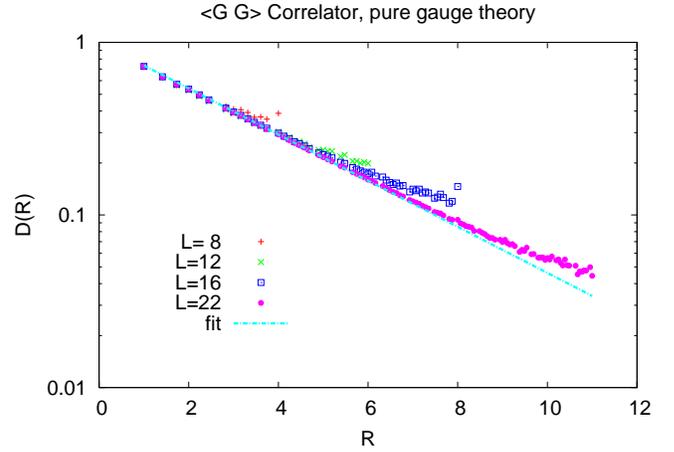}}}
\caption{The gauge transformation correlator $D(R)$ vs.\ $R$.  The correlator is evaluated, at various $L^4$ lattice volumes,
for configurations $U^{(1)}$ generated by lattice Monte Carlo in SU(2) pure gauge theory at $\b=2.2$, 
and $U^{(2)}$ derived from  $U^{(1)}$ with noise parameter $\kappa=5.67$ ($\oh \overline{\text{Tr}(\r)}=0.75$).
The straight line is a fit of $\exp(-\mu R)$ to the data at low $R$. }
\label{DR}
\end{figure}

    We begin by generating, via lattice Monte Carlo simulations, a set of thermalized SU(2) lattices, using the usual
Wilson action in $D=4$ dimensions at $\b=2.2$.   Take any time slice of such a lattice, and denote the link variables as $U^{(1)}_k(\bx)$.
The configuration is fixed to Coulomb gauge by the over-relaxation method, and the gauge transformation (a product of the transformations 
obtained at each over-relaxation sweep) taking $U^{(1)}$ to Coulomb gauge is denoted $g_1(\bx)$.  Next we construct a ``nearby" lattice
$U^{(2)}_k(\bx)$ by adding a small amount of noise to each link variable in the original (non-gauge fixed) $U^{(1)}$, i.e.
\beq
U^{(2)}_k(\bx) = \r_k(\bx) U^{(1)}_k(\bx)
\eeq
where $\r_k(\bx)$ is a stochastic SU(2)-valued ``noise" field, biased towards the identity, and generated independently at each link
with probability distribution
\beq
            \text{prob.\ measure} \propto \exp\Bigl[\k \oh \text{Tr}(\r) \Bigr] d\r
\label{P}
\eeq
where $d\r$ is the Haar measure.
With this probability measure, the average value of $\text{Tr}(\r)$ as a function of $\k$ is
\beq
            \overline{\text{Tr}[\r]} =  {I_2(\k) \over I_1(\k) }
\eeq
Having generated $U^{(2)}$ in this way, we fix it to Coulomb gauge by the same over-relaxation procedure that was
applied to $U^{(1)}$, and obtain $g_2(\bx)$.   From this we construct 
$G(\bx)=g_2(\bx) g^\dg_1(\bx)$, and obtain, on a lattice of extension $L$ in the spatial directions, the correlator
\beq
            D(R) = {1\over L^3} \sum_{\bx} \text{Tr}[G(\bx) G^\dg(\bx + {\mathbf R})]
\eeq
The final step is to average the values of $D(R)$ obtained on every time-slice of every lattice of a set of independent, thermalized lattices.
The result, for average Tr($\r$)=0.75 ($\k=5.67$) and a variety of $L^4$ lattice volumes, is shown in Fig.\ \ref{DR}.  It is clear that $D(R)$ does indeed 
fall off exponentially, with inverse correlation length $\m=0.30$.

\begin{figure}[h!]
\centerline{\scalebox{0.7}{\includegraphics{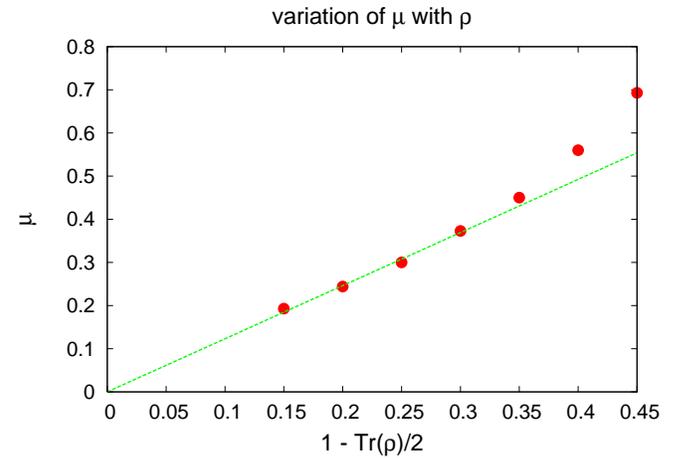}}}
\caption{Inverse correlation length $\m$, extracted from $D(R)$ calculated for various
values of the noise parameter $\kappa$, corresponding to $\oh \text{Tr}(\r)$ in the range $0.55-0.85$.
Errorbars are smaller than the symbol size.  The straight line is a fit to the first four data points.}
\label{rho}
\end{figure}

    As $\k \ra \infty$ and $\text{Tr}(\r) \ra 1$, it must be that the inverse correlation length $\m$ goes to zero. Fig.\ \ref{rho} is a plot of $\m$ vs.\
average $\text{Tr}(\r)$.  For lattice configurations $U^{(1)}$ generated (without gauge fixing)
at $\b=2.2$, the results are consistent with
\beq
            \m = c \Bigl( 1 - \oh \overline{\text{Tr}[\r]} \Bigr)
\eeq
and $c = 1.23$, as the average $\text{Tr}[\r]\ra 1$.   On a finite lattice (the maximum size used here is $L^4=22^4$), the practical constraint
on $\k$ is that it should not be so large that finite size effects are dominant.  Note also that $g_2\ra g_1$ only for $U^{(2)}$ approaching $U^{(1)}$, rather than approaching some arbitrary gauge copy of $U^{(1)}$.  If, e.g., $U^{(2)}$ is taken to be a random gauge copy of $U^{(1)}$, and the $U^{(1,2)}$ are transformed via over-relaxation to Coulomb gauge, then the gauge-fixed configurations are, in general, Gribov copies of one another.   Numerically it is found that $D(R)$ is consistent with zero in this case, for all $R>0$.  

\begin{figure}[t!]
\centerline{\scalebox{0.7}{\includegraphics{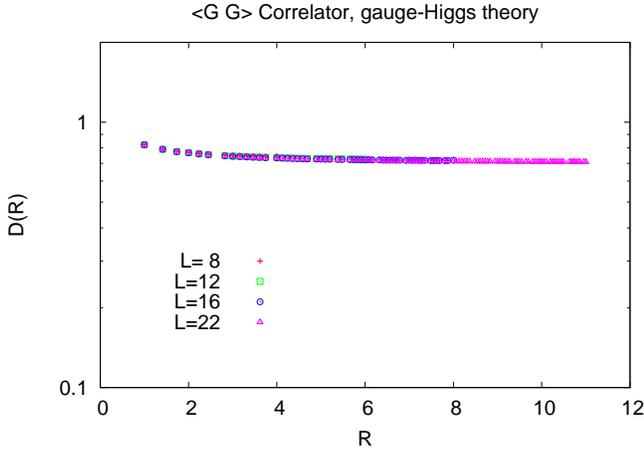}}}
\caption{Same as Fig.\ \ref{DR}, except that the $U^{(1)}$ are generated from lattice Monte Carlo simulation of
SU(2) gauge-Higgs theory (eq.\ \rf{higgsact}), in the Higgs-like region at couplings $\b=2.2,~\gamma=1.2$.  Noise
parameter $\kappa=5.67$ is the same as in Fig.\ \ref{DR}.}
\label{hprop}
\end{figure}

    Finally, since the conjecture is that $\langle G\rangle = 0$ results from a high density of near-zero F-P eigenvalues, as compared
to the density in a free field or non-confining theory,  we should find, conversely, that (i) $\langle G\rangle \ne 0$;  and (ii) $D(R)$ has a non-zero limit 
as $R\ra \infty$;  when $G(\bx)$ is evaluated for field 
configurations $U^{(1)}$ which do not have this high density of near-zero F-P eigenvalues.  More precisely, we expect $\langle G\rangle \ne 0$
for configurations in which the rhs of \rf{dth} is finite in the infinite volume limit.   Such configurations can be generated, e.g., by center
vortex removal in confining lattices, or alternatively by Monte Carlo simulation of an
SU(2) gauge-Higgs theory 
\beq
           S = \b \sum_{plaq} \oh \text{Tr}[U(P)] 
       + \gamma \sum_{x,\m} \oh \text{Tr}[\varphi^\dg(\bx) U_\m(\bx) \varphi(x+\widehat{\m})]  
\label{higgsact}
\eeq
in the ``Higgs-like" region of parameter space.  Here $U(P)$ denotes one-plaquette loops, and $\varphi(\bx)$ is an SU(2) matrix-valued Higgs field. 
It was shown in \cite{GOZ1} that the eigenvalue densities are qualitatively very similar in vortex-removed and gauge-Higgs configurations, and
are essentially a perturbation of the free field result.  The result for $D(R)$ in a gauge-Higgs theory at $\b=2.2,~ \g=1.2$ is shown in
Fig.\ \ref{hprop}.  Here we see that $D(R)$ does not have an exponential falloff; in fact it appears to approach a limiting value.   Hence $\m=0$
and $\langle G \rangle \ne 0$ in this case, as expected.

\section{\label{cc}Coulomb confinement}

      It is time to return to the question posed in the Introduction:  how do near-zero F-P eigenmodes enforce $\langle \text{Tr}[U_0]\rangle = 0$, which implies Coulomb confinement?  
       
     Consider a Monte Carlo simulation carried out in temporal gauge, with asymmetry parameter $\xi$.   The SU(2) Wilson action is given by
\bea
           S  &=&     {\b \over \xi} \sum_{\bx,t} \sum_{k=1}^3 \oh \text{Tr}[U_k(\bx,t) U^\dg_k(\bx,t+1)]
\non \\
                &+& \b \xi \sum_{\bx,t} \sum_{i<j} \oh \text{Tr}[U_i(\bx,t) U_j(\bx+\widehat{e}_i,t) 
                                       U^\dg_i(\bx+\widehat{e}_j,t) U^\dg_j(\bx,t)]
\non \\                                      
\eea
Suppose that $\xi \ll 1$.   In that case, the first term in the action will require that link variables $U_k(\bx,t)$ and $U_k(\bx,t+1)$
are almost identical.  This means that we may think of the sets of link variables at fixed times $t$ and $t+1$ as being an instance
of  ``nearby" $D=3$ dimensional lattice configurations $U^{(1)}$ and $U^{(2)}$.   In fact, writing
\beq
           U_k(\bx,t+1) = \eta_k(\bx,t) U_k(\bx,t)
\eeq
the lattice action has the form
\bea
           S  =     {\b \over \xi} \sum_{\bx,t} \sum_{k=1}^3 \oh \text{Tr}[\eta_k(\bx,t)]
                + \b \xi \sum_{\bx,t} \sum_{i<j} \oh \text{Tr}[U(P_{ij}(\bx,t))]
\non \\                                      
\eea
where $U(P_{ij}(\bx,t))$ is a loop around the space-like plaquette $P_{ij}(\bx,t)$.
Then, if $Q$ is any functional of the $\eta_k(\bx,t)$ at fixed $t$,
\begin{widetext}
\bea
   \langle Q \rangle
  &=&  \int DU(t+1) DU(t) ~ \Psi^*_0\Bigl[U(t+1)\Bigr] Q[\eta(t)] 
     \exp\left[ {\b \over \xi} \sum_{\bx} \sum_{k=1}^3 \oh \text{Tr}[\eta_k(\bx,t)] 
  + \xi \b \times \text{space-like plaquettes}  \right]  \Psi_0\Bigl[U(t)\Bigr]
\non \\
  &=&  \int D\eta(t) DU(t) ~ \Psi_0^*\Bigl[\eta(t) U(t)\Bigr] \Psi_0\Bigl[U(t)\Bigr] Q[\eta(t)] \exp\left[ {\b \over \xi} \sum_{\bx} 
           \sum_{k=1}^3 \oh \text{Tr}[\eta_k(\bx,t)]
  + \xi \b \times \text{space-like plaquettes}  \right]
\eea
\end{widetext}
where $\Psi_0[U]$ is the ground state (i.e.\ lowest energy eigenstate of the transfer matrix) in temporal gauge.
For $\xi \ll 1$ we have, to leading order in $\xi$,
\beq
   \langle Q \rangle
    =  \int D\eta(t) ~ Q[\eta(t)] 
     \exp\left[ {\b \over \xi} \sum_{\bx} \sum_{k=1}^3 \oh \text{Tr}[\eta_k(\bx,t)] \right] 
\eeq
So to leading order there are no correlations between the $\eta_k(\bx,t)$ at different lattice sites,\footnote{For correlations among electric field operators at different sites, and also to compute the Hamiltonian operator, one must of course keep the subleading terms.}
and the probability measure for $\eta_k(\bx)$ at a given link $k,\bx$ is simply
\beq
     \text{prob. measure} \propto \exp\left[ {\b \over \xi} \oh \text{Tr}[\eta] \right] d\eta
\label{peta}
\eeq
where $d\eta$ is the Haar measure.
Identifying $\k = \b/\xi$, this is exactly the same as the probability measure for the noise variable $\r$ of the previous section, and for
$\xi \ll 1$ it is strongly biased towards the identity. Therefore, $U^{(1)}=\{U_k(\bx,t)\}$ and $U^{(2)}=\{U_k(\bx,t+1)\}$ are 
close together in the space of $D=3$ lattice configurations.

     Now to compute the Coulomb potential, we transform the temporal gauge lattice configurations to Coulomb gauge via, e.g., over-relaxation.
These transformations can be computed independently at each time slice.   Pick any time $t$, and  denote
\bea
           U^C_k(\bx,t)  &=& g_1(\bx) U_k(\bx,t) g_1^\dg(\bx + {\bf e}_k)
\non \\
           U^C_k(\bx,t+1)  &=& g_2(\bx) U_k(\bx,t+1) g_2^\dg(\bx + {\bf e}_k)
\eea
where the ``C" superscript indicates the link variables transformed to Coulomb gauge, with $g_1$ and $g_2$ the gauge transformations
which take the temporal gauge lattices, at times $t$ and $t+1$ respectively, into Coulomb gauge.  Then the timelike link variables at time
$t$ are simply a product of these tranformations, i.e.
\bea
           U^C_0(\bx,t) &=& g_1(\bx) g_2^\dg(\bx)
\non \\
                            &=& G^\dg(\bx)
\label{UC}
\eea
But this means that
\beq
          \langle \mbox{Tr}[U_0(\bx,t)] \rangle =   \langle \mbox{Tr}[G(\bx)] \rangle
\eeq
and also, referring back to eq.\ \rf{my_way}, that
\beq
            V_C(R) =  - \xi^{-1} \log \Bigl[ \langle \text{Tr}[G(\bx) G^\dg(\bx+{\bf R})] \rangle\Bigr] - \text{const.} 
\eeq
The conclusion is that the property $\langle G \rangle = 0$, which requires that Coulomb confinement condition {\bf A} (eq.\ \rf{A}) is satisfied, in turn implies Coulomb confinement condition {\bf B}, i.e.\ $\langle U^C_0 \rangle = 0$. The latter condition also tells us that remnant global gauge symmetry in Coulomb gauge, shown in eq.\ \rf{remnant}, is unbroken (cf.\ the discussion in ref.\ \cite{GOZ2} on this point).   Moreover, a finite correlation length among the $G(\bx)$ implies a linear Coulomb potential.

      In the notation of the previous section, $G(\bx) = \exp[i\d \th(\bx)]$, and from eq.\ \rf{UC}, we see that when $G(\bx)$ is obtained from temporal
gauge, $\d \th^a$ plays the role of $g \xi A_0^a(\bx)$.  
It must be stressed, however, that while we can obtain $U_0^C$ by exponentiating $\d \th = g\xi A_0$, we cannot obtain $\d \th$ by simply taking the logarithm of $U_0^C$.  There is an issue of which branch of the logarithm to choose, and choosing, e.g., the principal value prescription, so that $\d \th^a$ and $A_0^a$ run over a finite range, would mean that the $00$ component of the gluon propagator is strictly bounded.  In fact, in contrast to the
$U_0 U_0$ correlator, the lattice $\langle A_0 A_0\rangle$ correlator cannot possibly result in a confining Coulomb potential, as seen explicitly
in the Appendix.
 
   For calculation of the Coulomb potential, we are considering $U^{(1)}$ and $U^{(2)}$ as adjacent time-slices of a thermalized configuration in temporal gauge, whereas in the simulations of the previous section, $U^{(1)}$ is a time-slice of a thermalized lattice generated without any gauge fixing, and $U^{(2)}$ is obtained from $U^{(1)}$ by adding a little noise.   According to the geometric picture advocated above, this difference is unimportant so far as the finite correlation length is concerned. The crucial
property leading to a finite correlation length among the $G(\bx)$ is that  $U^{(1)}$ has a high density of near-zero F-P modes when transformed
to Coulomb gauge, and that $U^{(2)}$ is obtained by a small displacement from $U^{(1)}$ in field space, in a random direction.   

    Two last comments are in order, regarding the continuous time $\xi \ra 0$ limit.  First, from eqs.\ \rf{DR1} and \rf{UU}, we see that in units of the lattice spacing in the spatial directions,
\beq
             \m = \xi \s_C
\eeq
so if $\s_C$ is finite and non-zero in the continuous time limit, it must be that the inverse correlation length $\m$ is proportional to $\xi$
as $\xi \ra 0$.  Now, from the probability measure \rf{peta} for $\eta$, we have, for small $\xi/\b$,
\bea
           \langle \oh \text{Tr}[\eta] \rangle &=& {I_2(\b/\xi) \over I_1(\b/\xi) }
\non \\
                  &=& 1 - {3\over 2} {\xi \over \b} + O\left({ \xi^2 \over \b^2} \right) 
\eea
or
\beq
                \xi \approx {2\over 3} \b \Bigl(1 -  \langle \oh \text{Tr}[\eta] \rangle \Bigr)
\eeq
Therefore, if $\m \propto \xi$ as $\xi \ra 0$, it must also be true that $\m \propto 1 -  \langle \oh \text{Tr}[\eta] \rangle$ in the same limit.   This seems consistent
with the data in Fig.\ \ref{rho} of the previous section. 

     The second comment is that, since the timelike link correlation length $l=\m^{-1}$ runs to infinity in the continuous time limit, 
there will naturally be long range correlations in that limit between two timelike link variables, representing a static quark-antiquark pair, and a timelike
plaquette variable.  What this means is that the Coulomb electric field, although confining, is \emph{not} collimated into a flux tube.  This is
the ``dipole problem" associated with all models of confinement based on one-particle exchange forces.  In general such models are prone to 
long-range dipole fields associated with static charges, and long-range van der Waals forces among hadrons.  This problem may be solved,
or at least alleviated, in the framework of the gluon chain model (cf.\ ref.\ \cite{new}).  \\

\section{Conclusions}

     We have seen that disorder ($\langle \text{Tr}[U_0]\rangle = 0$) in the Coulomb gauge timelike link variables, which implies Coulomb
confinement, can be traced to the fact that nearby gauge configurations, when transformed to Coulomb gauge, wind up far apart in the space of lattice configurations.  This feature of gauge orbits requires that a certain condition on near-zero modes of the Fadeev-Popov operator is satisfied, and this is in fact the \emph{same} condition derived from requiring that Coulomb self-energy of an isolated color charge is infinite. 
 
     Faddeev-Popov near-zero modes can also be thought of as near-invariances of the gauge-fixing condition.  The occurrence of a great number of such near-zero modes means that, while the Coulomb gauge (in, say, the fundamental modular region) may indeed be a complete gauge-fixing condition, for typical lattices in a confining theory it is just barely so; i.e.\ there are many directions in the gauge orbit which lift the gauge-fixed
lattice only slightly away from the gauge-fixing hypersurface; the gauge orbits are almost tangential, in many directions, to the gauge-fixing hypersurface.  It is not entirely clear why typical gauge orbits in a pure gauge theory have this property, while typical gauge orbits in a gauge-Higgs theory do not.  Proximity of a gauge orbit to the Gribov horizon is no doubt necessary but it is not sufficient, since it is known, e.g., that \emph{any}  lattice configuration in Coulomb gauge lying entirely in an abelian or center subgroup of the gauge group
is on the Gribov horizon \cite{GOZ2}, but not all abelian and center configurations are Coulomb confining.   
While there have been some very interesting  studies relevant to the F-P eigenmode spectrum (see, in particular, refs.\ \cite{Maas} and \cite{Holdom}), and it is known that center vortex removal has a drastic effect \cite{GOZ1}, it is probably fair to say that the density of Faddeev-Popov near-zero modes found in confining and non-confining theories is not yet well understood.

\acknowledgments{This research was supported in part by the U.S.\ Department of Energy under Grant No.\ DE-FG03-92ER40711. }  

\appendix*
\section{The Ambiguous $\bf A_0 A_0$ Propagator}

    According to eq.\ \rf{my_way} above, the Coulomb field energy of a quark-antiquark state can be extracted from the logarithm of the timelike link-link correlator, and this correlator can be formally expressed in terms of the $A_0$ component of the gluon field
\beq
\langle \mbox{Tr}[U_0^\dg(\bx,t) U_0(\by,t)] \rangle = \left\langle \mbox{Tr}\left[e^{-i\xi A_0(\bx,t)} 
e^{i\xi A_0(\by,t)} \right] \right\rangle
\eeq
In the spirit of exponentiating ladder diagrams, it may be argued that the exponential falloff of the link-link correlator is due to a
confining gluon propagator, since the instantaneous part of the 00 component of the gluon propagator is thought 
to be proportional to the Coulomb potential \cite{Dan-Attilio} .  Therefore, it may be expected that
\bea
\G_{00}(R) &=& \langle \oh \text{Tr}[A_0(\bx,t) A_0(\by,t)] \rangle 
\non \\
&\sim& {\s_C \over \xi} R
\label{AA}
\eea
asymptotically.   This relation may be true, but the problem is that there is no way to verify it on the
lattice.  The reason is simple:  $A_0(\bx)$ is basically the logarithm of $U_0(\bx)$, but the logarithm
of $U_0(\bx)$ is not unique.  It is necessary to choose a branch of the logarithm, and we have
no way of knowing which is the correct branch to choose.  Choosing one particular branch (e.g.\ via a
principle value prescription) cannot possibly result in a propagator satisfying \rf{AA}, because $R$ is only
limited by the lattice size, while the magnitude of $A_0$, and likewise $A_0(\bx) A_0(\by)$, is strictly bounded.

\begin{figure}[b!]
\centerline{\scalebox{0.6}{\includegraphics{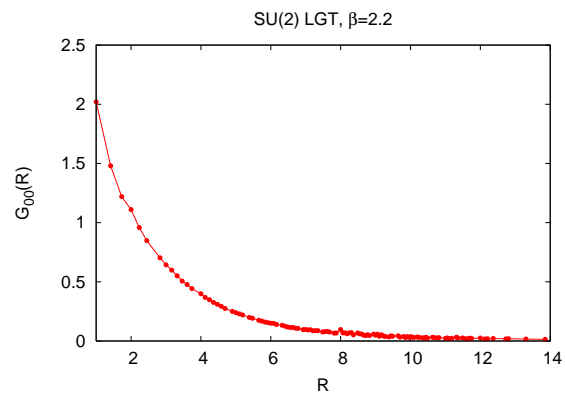}}}
\caption{Monte Carlo data for the $00$ component of the equal-times gluon propagator, with $A_0^a$ extracted from the timelike link
variable as described in the text.  There is no evidence of a confining Coulomb potential.  The simulation is carried out on a $16^4$ lattice in SU(2) lattice gauge theory at $\b=2.2$.}
\label{aprop}
\end{figure}

    The difficulty is quite clearly illustrated by an explicit Monte Carlo calculation of the lattice gluon propagator $\G_{00}(R)$ 
carried out at $\b=2.2,~\xi=1$ on a $16^4$ lattice.   To eliminate the multi-valuedness ambiguity, we extract $A_0$ from 
$U_0=\exp[i{\bf A}_0 \cdot {\mathbf \s} ]$ with the condition that $0 \le |{\bf A}_0| \le \pi$.   The result is shown in Fig.\ \ref{aprop}.  In contrast to the potential extracted from timelike link variables via eq.\ \rf{my_way} \cite{GO,japan}, there is no hint of a confining potential in the data for the gluon propagator $\G_{00}(R)$.

\begin{thebibliography}{}
\bibliographystyle{unsrt}

  

\bibitem{Dan}
  D.~Zwanziger,
  Phys.\ Rev.\ Lett.\  {\bf 90}, 102001 (2003)
  [arXiv:hep-lat/0209105].

\bibitem{Adam}
 A.~P.~Szczepaniak and P.~Krupinski,
  Phys.\ Rev.\  D {\bf 73}, 116002 (2006)
  [arXiv:hep-ph/0604098]; \\
 A.~P.~Szczepaniak and E.~S.~Swanson,
  Phys.\ Lett.\  B {\bf 577}, 61 (2003)
  [arXiv:hep-ph/0308268].; \\
  A.~P.~Szczepaniak and E.~S.~Swanson,
  Phys.\ Rev.\  D {\bf 65}, 025012 (2002)
  [arXiv:hep-ph/0107078]. 
   
\bibitem{gchain}
  J.~Greensite and C.~B.~Thorn,
  JHEP {\bf 0202}, 014 (2002)
  [arXiv:hep-ph/0112326].

\bibitem{new}
  J.~Greensite and \v{S}.\ Olejn{\'\i}k,
  arXiv:0901.0199 [hep-lat].


\bibitem{GOZ1}
  J.~Greensite, \v{S}.\ Olejn{\'\i}k, and D.~Zwanziger,
  JHEP {\bf 0505}, 070 (2005)
  [arXiv:hep-lat/0407032].

\bibitem{GOZ2}
  J.~Greensite, \v{S}.\ Olejn{\'\i}k and D.~Zwanziger,
  Phys.\ Rev.\  D {\bf 69}, 074506 (2004)
  [arXiv:hep-lat/0401003].

\bibitem{Parisi}
   E.~Marinari, M.~L.~Paciello, G.~Parisi and B.~Taglienti,
  Phys.\ Lett.\  B {\bf 298}, 400 (1993)
  [arXiv:hep-lat/9210021].


\bibitem{Dan1}
D.~Zwanziger,
  Nucl.\ Phys.\  B {\bf 518}, 237 (1998); \\
V.~Gribov, Nucl.\ Phys.\  B {\bf 139}, 1 (1978)

\bibitem{direct}
A.~Cucchieri and D.~Zwanziger,
  Nucl.\ Phys.\ Proc.\ Suppl.\  {\bf 119}, 727 (2003)
  [arXiv:hep-lat/0209068]; \\
K.~Langfeld and L.~Moyaerts,
  Phys.\ Rev.\  D {\bf 70}, 074507 (2004)
  [arXiv:hep-lat/0406024]; \\
 A.~Voigt, E.~M.~Ilgenfritz, M.~Muller-Preussker and A.~Sternbeck,
  Phys.\ Rev.\  D {\bf 78}, 014501 (2008)
  [arXiv:0803.2307 [hep-lat]].
 
 
\bibitem{GO}
  J.~Greensite and \v{S}.\ Olejn{\'\i}k,
  Phys.\ Rev.\  D {\bf 67}, 094503 (2003)
  [arXiv:hep-lat/0302018].

\bibitem{japan} 
  Y.~Nakagawa, A.~Nakamura, T.~Saito, H.~Toki and D.~Zwanziger,
  Phys.\ Rev.\  D {\bf 73}, 094504 (2006)
  [arXiv:hep-lat/0603010]; \\
  A.~Nakamura and T.~Saito,
  Prog.\ Theor.\ Phys.\  {\bf 115}, 189 (2006)
  [arXiv:hep-lat/0512042].


\bibitem{Mandula}
  J.~E.~Mandula and M.~Ogilvie,
  Phys.\ Lett.\  B {\bf 185}, 127 (1987).

\bibitem{Dan-Attilio}
  A.~Cucchieri and D.~Zwanziger,
  Phys.\ Rev.\  D {\bf 65}, 014002 (2002)
  [arXiv:hep-th/0008248].

  
\bibitem{Maas}
  A.~Maas,
  Eur.\ Phys.\ J.\  C {\bf 48}, 179 (2006)
  [arXiv:hep-th/0511307].

\bibitem{Holdom}
  B.~Holdom,
  arXiv:0901.0497 [hep-ph].


\end{thebibliography}
\end{document}